\documentclass[11pt,twocolumn,superscriptaddress]{revtex4}   %% LaTeX 2e (preferred)
\pdfoutput=1
\usepackage{amsmath}
\usepackage{graphicx}
\usepackage{natbib}
\usepackage{mathrsfs}

\usepackage[T1]{fontenc}
\usepackage[letterpaper,textwidth=7in,top=.75in,bottom=.75in]{geometry}
\begin{document}
\title{Temperature dependence of the nitrogen-vacancy magnetic resonance in diamond}

 \author{V.~M.~Acosta}
 \email{vmacosta@berkeley.edu}
    \address{
     Department of Physics, University of California,
     Berkeley, CA 94720-7300
    }

 \author{E.~Bauch}
    \address{
     Department of Physics, University of California,
     Berkeley, CA 94720-7300
    }
    \address{
     Technische Universit\"at Berlin,
     Hardenbergstra\ss e 28, 10623 Berlin, Germany
    }
 \author{M.~P.~Ledbetter}
    \address{
     Department of Physics, University of California,
     Berkeley, CA 94720-7300
    }
 \author{A.~Waxman}
    \address{Department of Physics, Ben-Gurion University,
    Be'er-Sheva, 84105, Israel
    }
 \author{L.-S.~Bouchard}
    \address{Department of Chemistry and Biochemistry, University of California,
    Los Angeles, CA 90095
    }
 \author{D.~Budker}
 \email{budker@berkeley.edu}
    \address{
     Department of Physics, University of California,
     Berkeley, CA 94720-7300
    }
    \address{Nuclear Science Division, Lawrence Berkeley National Laboratory,
     Berkeley CA 94720, USA
    }

\date{\today}

%\begin{article}
\begin{abstract}
The temperature dependence of the magnetic resonance spectra of nitrogen-vacancy (NV$^{\mbox{-}}$) ensembles in the range of $280\mbox{-}330~{\rm K}$ was studied. Four samples prepared under different conditions were analyzed with NV$^{\mbox{-}}$ concentrations ranging from $10~{\rm ppb}$ to $15~{\rm ppm}$. For all samples, the axial zero-field splitting (ZFS) parameter, $D$, was found to vary significantly with temperature, $T$, as $dD/dT=-74.2(7)~{\rm kHz/K}$. The transverse ZFS parameter, $E$, was non-zero (between 4 and $11~{\rm MHz}$) in all samples, and exhibited a temperature dependence of $dE/(E dT)=-1.4(3)\times10^{-4}~{\rm K^{-1}}$. The results might be accounted for by considering local thermal expansion. The temperature dependence of the ZFS parameters presents a significant challenge for diamond magnetometers and may ultimately limit their bandwidth and sensitivity.

%07.55.Ge Magnetometers for magnetic field measurements, 61.72.jn Color centers
%76.30.Mi Color centers and other defects
%81.05.Uw  Carbon, diamond, graphite
\end{abstract}
%\pacs{(07.55.Ge) Magnetometers for magnetic field measurements; (61.72.jn) Color centers; (76.30.Mi) Color centers and other defects; (81.05.Uw) Carbon, diamond, graphite}
\maketitle

Magnetometers based on nitrogen-vacancy (NV) ensembles in diamond \cite{TAY2008,ACO2009,BOU2010} promise high-sensitivity, rivaling those of superconducting quantum interference devices (SQUIDs) \cite{CLA2004} and alkali vapor magnetometers \cite{BUD2007}, in a scaleable solid state system that can be operated over a wide range of temperatures. This remarkable combination of spatial resolution \cite{BAL2008,HAN2009} and magnetic sensitivity \cite{BAL2009} make diamond magnetometers promising candidates for remote-detection and low-field nuclear magnetic resonance spectroscopy \cite{LEE2005,XU2006PNAS,SAV2007,LED2008PNAS}, nano-scale biological imaging \cite{MAZ2008NATURE,BAL2008,RUG2004,CLE2006}, and studies of novel magnetic and superconducting materials \cite{KIR1998,BOU2010}. Until now, the temperature dependence of the magnetic resonance spectra has not been systematically studied and has only briefly been mentioned in the literature \cite{GRU1997,BOU2010}. In this Letter, we report a striking temperature dependence of the magnetic-resonance spectra of NV$^{\mbox{-}}$ ensembles in diamond over the temperature range of $280\mbox{-}330~{\rm K}$. These findings have important implications for the design of diamond magnetometers and may ultimately limit their sensitivity.

The resonance spectra were recorded using the continuous-wave Fluorescence Detected Magnetic Resonance (FDMR) method \cite{HE1993A,GRU1997}. Light from a $514\mbox{-}{\rm nm}$ Argon-ion laser was focused with a $2.5~{\rm cm}$ focal length lens onto the diamond samples, exciting the NV$^{\mbox{-}}$ centers' $^3A_2\rightarrow {^{3}E}$ optical transition via a phonon sideband \cite{MAN2006}. The same lens was used to collect fluorescence from the diamond which was then passed through a dichroic mirror and a $650\mbox{-}800~{\rm nm}$ bandpass filter and detected with a photodiode. Noise due to laser power fluctuations was reduced by normalizing the fluorescence signal to a reference photodiode which monitored the incident laser power. The output of a microwave signal generator was amplified, passed through a straight $\sim200\mbox{-}{\rm \mu m}$ diameter copper wire of length $\sim5~{\rm mm}$ placed within $500~{\rm \mu m}$ of the focused light beam, and terminated with $50\mbox{-}{\Omega}$ impedance. For temperature control, the diamond was thermally connected to a copper heatsink and placed inside an insulated aluminum housing. The temperatures of the heat sink and housing were controlled with separate thermoelectric (TE) elements. Unless otherwise stated, the results reported in this Letter were obtained with a magnetic field of $\lesssim1~{\rm G}$, laser-light power of $\sim150~{\rm mW}$, and microwave power (after the wire) of $\sim10~{\rm dBm}$.

For temperature scans, the temperature of the copper plate in direct thermal contact with the diamond was monitored with an AD590 sensor. The FDMR spectra were recorded with the temperature stabilized so that temperature excursions were less than $0.05~{\rm K}$ over 5 min.  In order to avoid stray magnetic fields when recording the spectrum, the currents supplied to both TE elements were chopped at a frequency of $2~{\rm Hz}$ using photoMOS circuits, and the spectra were recorded only when the TE currents were off. The process was repeated until the temperature had been scanned through the $280\mbox{-}330~{\rm K}$ range several times in both directions.

The NV-ensemble magnetic-resonance spectroscopy has been described, for example, in Refs. \cite{LOU1978,HE1993A,ALE2007,LAI2009,ACO2009}) and is only briefly summarized here. Optical pumping via a spin-selective decay path collects NV centers (total spin $S=1$) in the $|m_s=0\rangle$ ground-state magnetic sublevel \cite{MAN2006}. In the absence of external fields, the $|m_s=0\rangle$ and $|m_s=\pm1\rangle$ levels are split by an energy equal to the axial zero-field splitting (ZFS) parameter, $D\approx2.87~{\rm GHz}$. For perfect $C_{3v}$ symmetry, the transverse ZFS parameter is $E=0$ and the $|m_s=\pm1\rangle$ levels remain degenerate. When the frequency of a microwave field that is transverse to the symmetry axis is tuned to the energy splitting between the $|m_s=0\rangle$ and $|m_s=\pm1\rangle$ levels, NV centers are transferred to the $|m_s=\pm1\rangle$ sublevels, resulting in diminished fluorescence with a contrast as high as $30\%$ \cite{BAL2008}. In the presence of an applied magnetic field, $B$, the $|m_s=\pm1\rangle$ levels split, revealing resonances separated by $2 g_{NV} \mu_B B$, where $g_{NV}=2.003$ is the NV$^{\mbox{-}}$ Land\'e factor \cite{LOU1978,FEL2009} and $\mu_B$ is the Bohr magneton. For ensembles, there are four different NV orientations and, provided that $g_{NV}\mu_B |B|\ll D$, only the projection of the magnetic field on the N-V axis affects the transition frequencies \cite{LAI2009}.

The zero-field Hamiltonian for the ground state, including hyperfine coupling to the $^{14}$N nucleus (spin $I=1$), can be written as:
\begin{equation}
\label{eq:hamiltonian}
\begin{split}
\mathscr{H}_0\approx D S_z^2+E (S_x^2-S_y^2)%+P I_z^2
\\
+A_{\parallel} S_z I_z+A_{\perp}(S_x I_x+S_y I_y),
\end{split}
\end{equation}
where $A_{\parallel}=-2.1~{\rm MHz}$ and $A_{\perp}=-2.7~{\rm MHz}$ are, respectively, the axial and transverse hyperfine constants \cite{FEL2009}. Analysis of this Hamiltonian reveals six allowed microwave transitions for each N-V orientation. The relative intensities can be calculated by treating the interaction with the microwave field, $\vec{B_1}$ as a perturbation, $\mathscr{H}_1=g_{NV}\mu_B\vec{B_1}\cdot\vec{S}$, with matrix elements that depend on the alignment of the microwave radiation with respect to the symmetry and strain axes of each N-V center. However, since the exact geometry and the number of NV$^{\mbox{-}}$ centers of each orientation were not known \emph{a priori}, Gaussian functions with variable amplitudes and equal widths, centered about these transition frequencies, were fit to the spectra. Including residual magnetic fields, measured by a commercial fluxgate magnetometer to be less than $1~{\rm G}$, into the model did not significantly influence the fits.

Four single-crystal samples of mm-scale dimensions were studied, which were labeled S2, S3, S5, and S8 and characterized in Ref. \cite{ACO2009}.
\begin{figure}
\centering
    \includegraphics[width=.45\textwidth]{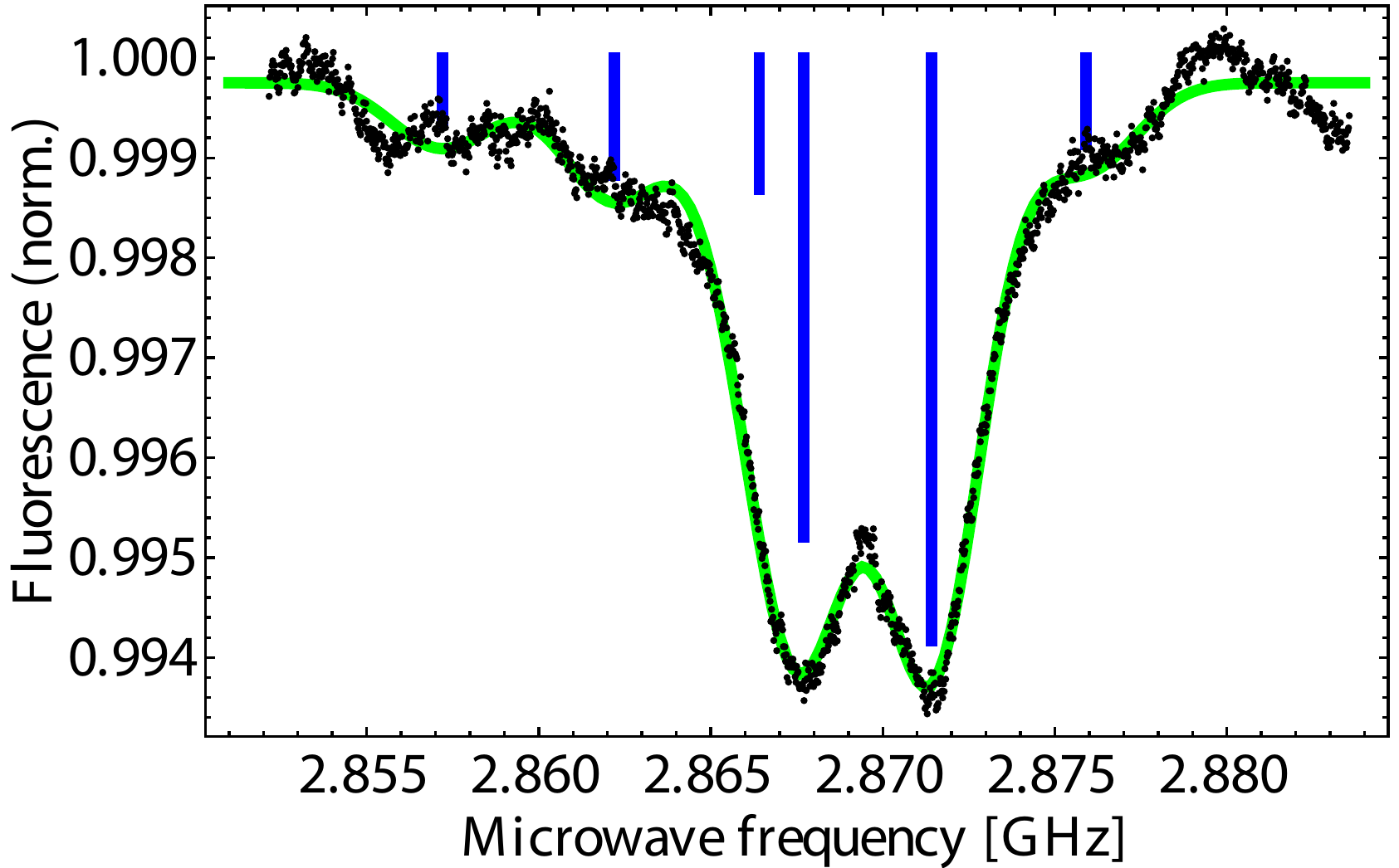}
    \caption{\label{fig:raws3} Zero-field FDMR spectrum at $293~{\rm K}$ for S3 and the corresponding fit based on Eq. \ref{eq:hamiltonian} (solid green line). The six blue lines represent the fitted amplitudes at each transition frequency, and the fitted linewidth was $3.3~{\rm MHz}$ (full width at half maximum). The microwave power was reduced to $\sim-10~{\rm dBm}$ to resolve the hyperfine structure, resulting in the relatively small contrast of $\sim0.6\%$. The best-fit parameters for this scan are $E=4.1(2)~{\rm MHz}$ and $D=2866.8(2)~{\rm MHz}$. }
\end{figure}
Figure \ref{fig:raws3} shows the FDMR spectrum at $293~{\rm K}$ for S3, a sample synthesized by chemical vapor deposition (CVD) with $[{\rm NV}^{\mbox{-}}]^-\approx10~{\rm ppb}$ \cite{ACO2009}. As there was no applied magnetic field, the splitting between resonance peaks is due to non-zero $E$, induced by local strain \cite{BAL2008,FUC2008,LAI2009}. This feature is present in varying magnitudes for all four samples. Even though all four NV orientations are present, the spectra are reasonably well-described by just six broad transitions, suggesting that the strain splittings are spatially inhomogenous \cite{GRU1997, NIZ2001}. As no correlation with NV$^{\mbox{-}}$ concentration was observed (see Tab. \ref{tab:values}), further work is necessary to determine the exact strain mechanism.

During each temperature scan, the spectrum was fit to an empirical function similar to the one described above, and the ZFS parameters were extracted.
\begin{figure}
\centering
    \includegraphics[width=.45\textwidth]{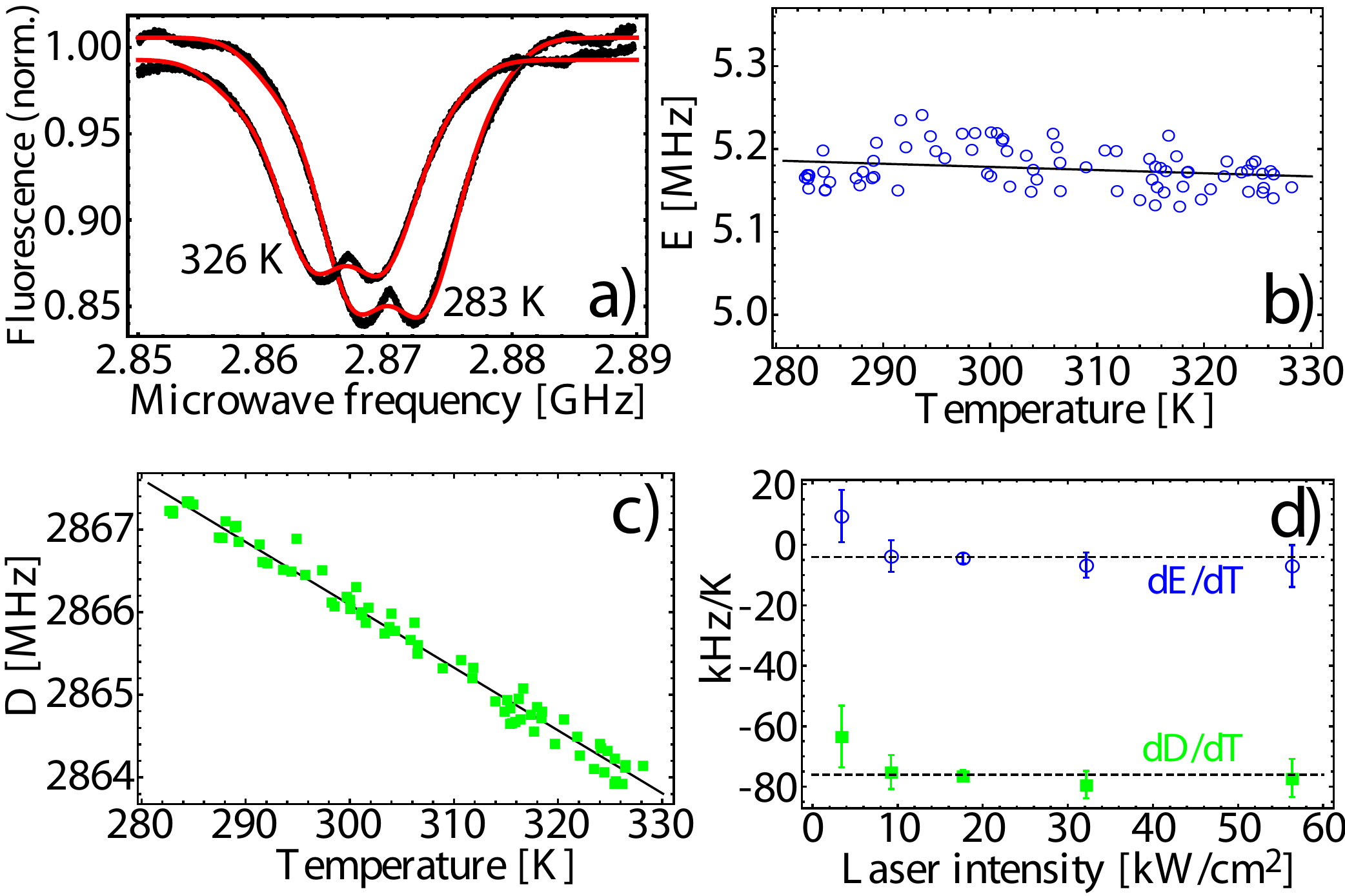}
    \caption{\label{fig:tempfit}(a) Zero-field FDMR spectra at $283~{\rm K}$ and $326~{\rm K}$ for S8 with fits (solid red lines). (b) Value of $E$ for S8 as a function of temperature with linear fit (solid black line). (c) $D$ for S8 vs. temperature with linear fit. (d) $dD/dT$ and $dE/dT$ as a function of laser intensity for S5. The dotted lines are the laser-intensity-independent values used in Tab. \ref{tab:values}. }
\end{figure}
Figure \ref{fig:tempfit}(a) displays the spectra at two different temperatures for another sample, S8, a high-pressure, high-temperature (HPHT) synthesized diamond with $[{\rm NV}^{\mbox{-}}]\approx0.3~{\rm ppm}$, as well as the empirical fits based on Eq. \ref{eq:hamiltonian}. Figures \ref{fig:tempfit}(b) and (c) show the ZFS parameters as a function of temperature for this sample. Linear least-squares fits yield $dE/dT=-0.4(2)~{\rm kHz/K}$ and $dD/dT=-76(1)~{\rm kHz/K}$. Figure \ref{fig:tempfit}(d) displays the laser-intensity dependence for S5, an HPHT diamond with $[{\rm NV}^{\mbox{-}}]\approx12~{\rm ppm}$. Linear fits (not shown) determined that any dependence of $dD/dT$ or $dE/dT$ on laser intensity is not statistically significant. Additional tests for dependence on microwave power, external magnetic field, and sample positioning also did not show statistically significant effects.

A similar procedure was performed for the three other samples: S2, an HPHT diamond with $[{\rm NV}^{\mbox{-}}]\approx16~{\rm ppm}$, as well as S3 and S5 (already mentioned).
\begin{table}[ht]
\centering
    \begin{tabular}{c c c c c c}
      \hline
      \hline
      \# & $[{\rm NV}^{\mbox{-}}]{\rm (ppm)}$ & ~$\frac{dD}{dT}({\rm kHz/K})$ & ~$E({\rm MHz})$ & ~$\frac{1}{E}\frac{dE}{dT}(10^{-4}{\rm K^{-1}})$~\\
      \hline
      S2 & $16$ & $-71(1)$ & $5.8(3)$ & $-1.7(5)$ \\ %\hline
      S3 & $0.01$ & $-79(2)$ & $4.3(2)$ & $2(5)$ \\ %\hline
      S5 & $12$ & $-77(3)$ & $11(1)$ & $-3.6(9)$\\ %\hline
      S8 & $0.3$ & $-76(1)$ & $5.2(1)$ & $-0.8(4)$ \\ %\hline
      %dE/dT (kHz/K) for {S2,S3,S5,S8} = {-1.0(3),1(2),-4(1),-0.4(2)}
      [1ex]
      \hline
    \end{tabular}
    \caption{\label{tab:values} ZFS parameters and uncertainties for four different samples. The values of $E$ represent the expected value of $E(293~{\rm K})$ extrapolated from the linear fits, and the error bars represent the standard error from the fit but not systematic effects due to imperfect assumptions in the model (see text). The laser intensity was $\sim25\mbox{-}50~{\rm kW/cm^2}$ throughout the collection volume. Note that for the S2 spectra a magnetic field of $B_{\perp}\approx13~{\rm G}$ was applied. This field enabled the isolation of a single NV orientation, and the simplified spectrum was used to verify the robustness of the model.}
\end{table}
Table \ref{tab:values} displays the temperature dependence of the ZFS parameters for each of these samples. The temperature dependence of $D$ is similar for each sample, indicating that the mechanism responsible for this temperature variation is intrinsic to the NV centers themselves. Taking a weighted average over all samples gives $dD/dT=-74.2(7)~{\rm kHz/K}$ and, using the fitted room-temperature values for each sample ($D\approx2867(1)~{\rm MHz}$), this corresponds to a fractional temperature dependence of $dD/(D dT)=-2.59(2)\times10^{-5}~{\rm K^{-1}}$. The weighted average over samples of the fractional variation of $E$ with temperature (final column of Tab. \ref{tab:values}) is also statistically significant, $dE/(E dT)=-1.4(3)\times10^{-4}~{\rm K^{-1}}$, but further work is necessary to understand the nature of $E$.

% D(293 K) for {S2,S3,S5,S8} in MHz is: {2866.17,2867.83,2869.07,2866.73)
The origin of $D$ is expected to be predominately due to dipolar spin-spin coupling between the two unpaired electrons forming the center \cite{LOU1978,LEN1996,HE1993A}. This suggests a a likely mechanism for the temperature variation is local lattice expansion. Assuming that the angular electronic wavefunctions are temperature-independent and that $D$ is entirely due to dipolar coupling, the effect of lattice expansion on $D$ is:
\begin{equation}
\label{eq:dR}
 \frac{1}{D}\frac{dD}{dT}\approx\frac{1}{D}\frac{d\langle (r_{12}^2-3z_{12}^2)/r_{12}^5 \rangle}{dR}\frac{dR}{dT},
\end{equation}
 where $r_{12}$ is the displacement between the two spins, $z_{12}$ is the component of $r_{12}$ along the N-V symmetry axis, and $R$ is the distance between two basal carbon nuclei. The effect of thermal expansion on $\langle (r_{12}^2-3z_{12}^2)/r_{12}^5 \rangle$ can be estimated by treating spins, localized near the basal carbon atoms \cite{HE1993A,GAL2008}, with $p$-orbitals \cite{HE1993A,FEL2009} oriented along axes $110^{\circ}$ apart \cite{GOS1996}, and calculating the integral for neighboring values of $R$. Using the room-temperature values for bulk diamond of $R=0.252~{\rm nm}$ and $dR/dT=2.52\times10^{-5}~{\rm nm/K}$ \cite{SAT2002}, we calculate $D=2.66~{\rm GHz}$, which is within $10\%$ of the experimental value, and $dD/(D dT)=-5.8\times10^{-6}~{\rm K^{-1}}$, which is about a factor of $4.5$ smaller than the experimental value from this work. The latter discrepancy suggests that the macroscopic thermal expansion is not a good description of $dR/dT$ in the immediate vicinity of the defect. \emph{Ab initio} calculations \cite{LEN1996,GOS1996,LUS2004,GAL2008,GAL2009} which include the determination of local thermal expansion effects would give a more accurate prediction of $dD/dT$.

The sharp temperature dependence of $D$ presents a technical challenge for room-temperature diamond magnetometry. Even if the ambient temperature can be controlled at the $1\mbox{-}{\rm mK}$ level, this would lead to fluctuations in the resonance frequency of $80~{\rm Hz}$ corresponding to a magnetic-field variation of $3~{\rm nT}$. Monitoring both of the $\Delta m_s=\pm1$ resonances could provide a feedback mechanism for controlling this effect for slow drifts, since the energy difference between these resonances does not depend on $D$.

Higher-frequency temperature fluctuations due to, for example, laser-intensity noise, present an additional complication for magnetometry in the high-density limit. Consider the case of a Ramsey-type magnetometer making use of repeated light pulses \cite{TAY2008,MAZ2008NATURE,BAL2008,BAL2009} which transfer an energy to the diamond on the order of $\displaystyle E_p\approx\Delta\epsilon[{\rm NV}^{\mbox{-}}] V$, where $\displaystyle \Delta\epsilon\approx0.6~{\rm eV}$ is the difference in energy between absorbed and radiated photons, $V$ is the effective volume being heated, and we have conservatively neglected non-radiative transfer from the NV$^{\mbox{-}}$ singlet decay path \cite{ROG2008} and other impurities \cite{DAV1973}. If the pulses are separated in time by a precession window, $\tau$, then in steady state the diamond temperature is modulated at a rate $\displaystyle \frac{dT}{dt}\approx \frac{E_p}{V c \tau}$, where $c=1.8~{\rm J/cm^3/K}$ is the volumetric specific heat of diamond \cite{VIC1962}. Integration over the precession window yields a magnetometer offset of $\displaystyle B_{off}\approx\frac{\pi\Delta\epsilon[{\rm NV}^{\mbox{-}}]}{g_{NV}\mu_B c} \frac{dD}{dT}\approx-80~{\rm nT}$ at room temperature for $[{\rm NV}^{\mbox{-}}]=1~{\rm ppm}$. This offset makes the magnetometer sensitive to laser-pulse fluctuations. Uncorrelated, normally-distributed fluctuations in $E_p$ by a fraction $\chi$ produce magnetic field noise-per-unit-bandwidth at the level of $\displaystyle \chi |B_{off}|/\sqrt{\tau}\approx1~{\rm pT/\sqrt{Hz}}$, using $\chi=0.01$ and $\tau=1~{\rm \mu s}$. We note that this magnetometer noise is directly correlated with laser-intensity noise and therefore monitoring the incident laser intensity could significantly reduce this effect.

In this work, we have measured the temperature dependence of the ZFS parameters of four diamond samples covering a wide range of NV$^{\mbox{-}}$ concentrations. We have found a significant variation of the axial ZFS, $D$, with temperature and surmise that it is due to local thermal expansion. We also present evidence of a non-zero transverse ZFS, $E$, and measure a small fractional temperature dependence just above the experimental uncertainty. The results have a major impact on the performance of NV-ensemble magnetometers and may ultimately limit their sensitivity and bandwidth. We expect that proper feedback mechanisms, such as monitoring laser intensity fluctuations and observing both $\Delta m_s=1$ coherences simultaneously, will help to partially mitigate these effects.

The authors are grateful to A. Gali, C. Santori, P. Hemmer, F. Jelezko, E. Corsini, and O. Sushkov for valuable discussions and R. Folman for support. This work was supported by NSF grant PHY-0855552 and ONR-MURI.

%\bibliography{diamond}

\end{document}